\documentclass[a4paper,twoside,12pt]{article}
\usepackage[T1]{fontenc}
\usepackage{amssymb}
\usepackage[latin1]{inputenc}
\usepackage{graphicx}

\begin{document}

\begin{center}
\Huge{\textbf{Yield to maturity modelling and a Monte Carlo Technique for pricing Derivatives on Constant Maturity Treasury (CMT) and Derivatives on forward Bonds}}\\
\vspace*{0.5cm}
\small{By}\\
\vspace*{0.5cm}
\large{\textbf{Didier KOUOKAP YOUMBI\footnote{Didier KOUOKAP YOUMBI(didier.kouokap@gmail.com) works at Societe Generale. Any opinions expressed here are those of the author and not necessarily those of the Societe Generale Group.\\
Special Thanks to Henri LEOWSKI and Hamid SKOUTTI for useful suggestions.}}}\\
%He is graduated from the Ecole Polytechnique in France. He is a former student of the master 2 Probabilities and Finance at Paris 6.\\Any opinions expressed here are those of the author and not necessarily those of the Societe Generale Group.}}}\\ 
\normalsize{First version: 29/02/2012. This version: March 30th, 2012}\\
\vspace*{0.5cm}
\vspace*{1cm}
\begin{abstract}
This paper proposes a Monte Carlo technique for pricing the forward yield to maturity, when the volatility of the zero-coupon bond is known. We make the assumption of deterministic default intensity (Hazard Rate Function). We make no assumption on the volatility of the yield. We actually calculate the initial value of the forward yield, we calculate the volatility of the yield, and we write the diffusion of the yield.

As direct application we price options on Constant Maturity Treasury (CMT) in the Hull and White Model for the short interest rate. Tests results with Caps and Floors on 10 years constant maturity treasury (CMT10) are satisfactory. This work can also be used for pricing options on bonds or forward bonds.

\end{abstract}
\end{center}

\newpage
\section*{Introduction}
The way most practitioners use to price CMT is to consider the CMT as a simple function of the CMS. Then the function's parameters are calibrated with the spreads between the forward CMS and the forward CMT. This spread has been increasing from beginning 2010, until mid-2011 when it stopped quoting. See Figure \ref{fig:histSpread} below.

\begin{figure}[htbp]
	\centering
		\includegraphics[height = 7cm, width = 13cm]{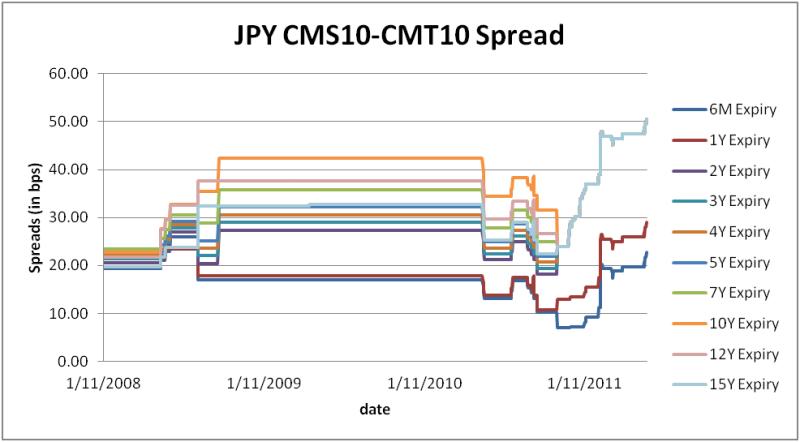}
	\caption{\textit{historical quotations of the spread between the JPY forward CMS10 and the JPY forward CMT10. This figure shows the explosion of the spread from the beginning 2010}}
	\label{fig:histSpread}
\end{figure}

But now there is no more liquidity for these spreads. As a consequence it has become very difficult to consistently price CMT and options on CMT, even with the inconsistent method described in the lines above.

The difficulty for pricing CMT properly relies in the ability of modelling the forward yield to maturity of the related bond. The dynamics of the forward yield is known. BENHAMOU (2000) writes the dynamics of the yield. However he did not propose how to calculate or calibrate the volatility of the yield. In this paper we make no assumption on the volatility of the yield. We actually compute it as well as we compute the volatility of the bond and the survival probability distribution.

In the first section of this paper we give notations, and we remind some definitions. In section 2 we write the dynamics of the forward yield to maturity, through the mathematical relation between the yield and the Bond price.  In the third section we use martingale condition to derive a partial derivative equation (pde) of which the hazard rate function is the solution. After solving the pde, we easily compute the volatility of the bond, then the volatility of the yield. In section 4 we make the assumption of a constant hazard rate function (with respect to the time between the forward start date and the maturity date), and we propose a Monte Carlo routine for pricing the expectation of the terminal yield to maturity, and the expectation of any payoff on the yield to maturity. Finally in section 5, in the hypothesis of a Hull and White model for the short rate, we give some tests results on the JPY 10Y maturity constant Treasury (CMT10), and options (Caplets, floorlets, Caps, Floors) on the CMT10. We draw the distribution of the CMT10 and the distribution of the volatility of the CMT10 for various expiries. We also compute the Black implied volatility related to the prices of Caps and Floors on the CMT10. As extension, we notice that this framework could be used for pricing options on bonds, or forward bonds, without further deve\-lopments.

\newpage

\section{Notations and Definitions}
In this section we give notations and we recall some definitions.

\subsection{Notations}
\begin{itemize}
	\item{	$B_{t,T}=Bond(t,T,T+\theta)$ : is the value at time t, of the T-forward $\theta$-Years constant maturity Bond price. We will refer to T as expiry, and T+$\theta$ will be the maturity;}
	\item{$y_{t,T}$ : is the T-forward $\theta$-Years constant maturity yield to maturity, related to the previous bond;}
	\item{	$c_i$ : is the value of coupon, expressed as a percentage of the notional, paid by the bond at time $T_i$;}
	\item{ $\tilde{c}$: is the coupon rate such that the value the coupon is equal to the coupon rate times the time step between the last payment date and the now payment date: $c_{i}=\tilde{c}(T_{i}-T_{i-1})$;}
	\item{$CMT(T,T+ \theta)$ is the value, at time T, of $\tilde{c}$   such that the value of the bond at time T is at par: $B_{T,T}=1$;}
	\item{$\kappa$ : is the number of times coupons are paid by the bond per year;}
	\item{$R$ : is the recovery rate of the bond issuer;}
	\item{$DF(t,T,T_{i})$: is the value at time t of the T-forward $T_{i}-T$ maturity discount factor: $DF(t,T,T_i)=\frac{DF(t,T_i)}{DF(t,T)}$;}
	\item{$P_{t,T,U}$ : is the value at time t of the T-forward  U-T maturity Zero-coupon Bond: $P_{t,T,U}=\frac{P_{t,U}}{P_{t,T}}$ 
}
	\item{$S(t,T,T_{i})$ : is the T-forward  $T_{i}-T$ maturity survival probability}
	\item{$\tau(t)$: is the first time after the time t, when the bond is subjected to a credit event;}
	\item{$\lambda(t,T,U)$ : is the T-forward  U-T maturity default intensity, also called hazard rate function. It will be properly defined in the next subsection.}
\end{itemize}

\subsection{Some Recalls}

\begin{itemize}
	\item{\textbf{Relation between the bond and the yield to maturity}}
\end{itemize}

\begin{eqnarray}
B_{t,T} &=& \sum_{i=1}^{{\kappa}{\theta}}{\frac{c_i}{(1+y_{t,T})^{\frac{i}{\kappa}}}} +\frac{1}{(1+y_{t,T})^{\theta}}\nonumber \\
 &=& f(y_{t,T}) \label{eq:Bond} 
\end{eqnarray}

With
$$f: x \longmapsto \sum_{i=1}^{{\kappa}{\theta}}{\frac{c_i}{(1+x)^{\frac{i}{\kappa}}}} +(1+x)^{-\theta}$$
And
$$x \in \mathbb{R}_{+}^{*}$$

In particular for constant coupons: $c_{i}=c \mbox{   }   \forall  \mbox{   }i \in \left\langle 1...{\kappa}{\theta}\right\rangle$, we have

$$f(x)=c\frac{1-(1+x)^{-\theta}}{(1+x)^{\frac{1}{\kappa}}-1}+(1+x)^{-\theta}$$

The inverse of the function f above will be denoted g:
$$g = f^{-1}$$

\begin{itemize}
	\item{\textbf{The Constant Maturity Treasury (CMT)}}
\end{itemize}
It is the value of the coupon rate $(\tilde{c})$ such that the Bond is at par, on the expiry date: $B_{T,T}=1$. It is always defined with a constant rolling $\theta$-maturity Bond.
Using the fact that ${\Delta}T_{i} \cong \frac{1}{\kappa}$; from equation \textbf{(\ref{eq:Bond})} we get that  $\tilde{c}$ , which is here equal to the CMT, should be the solution of the following equation:

$$1=\frac{\tilde{c}}{\kappa}\frac{(1-(1+y_{T,T})^{-\theta})}{(1+y_{T,T})^{\frac{1}{\kappa}}-1}+(1+y_{T,T})^{-\theta}$$

We find that
\begin{eqnarray}
 CMT(T,T+\theta)&=&{\kappa}\left((1+y_{T,T})^{\frac{1}{\kappa}}-1\right) \label{eq:CMT}\\
								& \cong & y_{T,T}\nonumber
\end{eqnarray}

For the $CMT\theta$ modelling, the expression of the corresponding forward Bond price can be rewritten as follow

\begin{eqnarray}
B_{t,T} &=& \left((1+y_{T,T})^{\frac{1}{\kappa}}-1\right)\frac{1-(1+y_{t,T})^{-\theta}}{(1+y_{t,T})^{\frac{1}{\kappa}}-1}+(1+y_{t,T})^{-\theta}\nonumber \\
 &=& f(y_{t,T}) \label{eq:BondCMT} 
\end{eqnarray}

With
$$f: x \longmapsto \left((1+y_{T,T})^{\frac{1}{\kappa}}-1\right)\frac{1-(1+x)^{-\theta}}{(1+x)^{\frac{1}{\kappa}}-1}+(1+x)^{-\theta}$$
And
$$x \in \mathbb{R}_{+}^{*}$$

This last version of function $f$ will be used when pricing the $CMT\theta$ or options on the $CMT\theta$.

\begin{itemize}
	\item{\textbf{Hazard Rate Function (or default intensity)}}
\end{itemize}

The hazard rate function is defined as follow:
\begin{eqnarray}
 {\lambda}(t,T)&=& {\lim_{{\Delta}T\rightarrow 0^{+}}}{\frac{\mathbb{P}_{t}(T<{\tau}(t)\leq T+{\Delta}T/{\tau}(t)>T)}{{\Delta}T}}\nonumber\\
								& = & {\lim_{{\Delta}T\rightarrow 0^{+}}}{\frac{\mathbb{P}_{t}(T<{\tau}(t)\leq T+{\Delta}T)}{{\Delta}{T\mbox{ }}\mathbb{P}_{t}({\tau}(t)>T)}} \nonumber\\
								& = & -{\lim_{{\Delta}T\rightarrow 0^{+}}}{\frac{S(t,T+{\Delta}T)-S(t,T)}{{\Delta}{T\mbox{ }}S(t,T)}} \nonumber\\
								& = & -\frac{1}{S(t,T)}\frac{{\partial}_{2}S}{{\partial}v}(t,T) \nonumber\\
								& = & -\frac{{\partial}_{2}{\ln}{S}}{{\partial}v}(t,T) \nonumber
\end{eqnarray}

And given the fact that $S(t,t)=1$ : no default has occured at initial time, we get that

$$S(t,T)= e^{-{\int}_{t}^{T}{{\lambda}(t,v)dv}}$$

Similarly, we express the forward hazard rate as follow

\begin{eqnarray}
 {\lambda}(t,T,U)&=& \mathbb{E}_{t}^{Q_{T}}\left[{\lambda}(T,U)\right]\nonumber\\
								& = & -\frac{{\partial}_{3}{\ln}{S}}{{\partial}u}(t,T,U) \nonumber
\end{eqnarray}

And
$$S(t,T,U)= e^{-{\int}_{T}^{U}{{\lambda}(t,T,u)du}}$$

\begin{itemize}
\item{\textbf{Zero-coupon Bond}}\\\\
The zero-coupon bond is the value of a contract that pays 1 at maturity. Under the risk-neutral measure, the dynamics of the zero-coupon Bond can be written as following:

$$\frac{dP_{t,T}}{P_{t,T}}=r_{t}dt+{\sigma}_{P}(t,T)dW_{t}$$

\begin{itemize}
	\item{${\sigma}_{P}(t,T)$ is the volatility of the zero coupon bond. In this framework we will suppose it to be known;}
	\item{$P_{T,T}=1$}
\end{itemize}

The T-Forward zero coupon Bond of maturity U is defined as follow
$$P_{t,T,U}=\frac{P_{t,U}}{P_{t,T}}$$

Then we have 
\begin{eqnarray}
\frac{dP_{t,T,U}}{P_{t,T,U}}={\xi}(t,T,U)dt+{\sigma}_{P}(t,T,U)dW_{t} \label{eq:ZCBondFwrd}
\end{eqnarray}

With
$${\xi}(t,T,U)={\sigma}_{P}(t,T)^{2} - {\sigma}_{P}(t,T){\sigma}_{P}(t,U)$$
And
$${\sigma}_{P}(t,T,U) = {\sigma}_{P}(t,U)-{\sigma}_{P}(t,T)$$

\end{itemize}

\section{Yield To Maturity dynamics}
Since the spot bond $(B_{t,t})$  is a tradable and replicable asset, it drifts at the risk-free rate under the risk neutral probability. Under this measure, the dynamics of the Bond can be written as follow:

$$\frac{dB_{t,t}}{B_{t,t}}=r_{t}dt+{\sigma}_{B}(t)dW_{t}$$

Where the process $(W_{t})_{(t \geq 0)}$ is a Brownian motion under the risk neutral probability.

And since $B_{t,T}$   is the forward, it should drift at zero, under the same measure: the forward is a local martingale under the risk neutral probability.

\begin{eqnarray}
\frac{dB_{t,T}}{B_{t,T}}={\sigma}_{B}(t,T)dW_{t} \label{eq:BondFwrd}
\end{eqnarray}

\textbf{Proposition 1}\\
\textit{The volatility of the forward yield can be expressed as a function of the volatility of the forward bond through the following formula}
\begin{eqnarray}
{\sigma}_{y}(t,T)=\frac{f(y_{t,T})}{y_{t,T}f^{'}(y_{t,T})}{\sigma}_{B}(t,T) \label{eq:VolYield}
\end{eqnarray}

\textit{The dynamics of the forward yield under the risk-neutral measure is the following}
\begin{eqnarray}
\frac{dy_{t,T}}{y_{t,T}}=-\frac{1}{2}\frac{y_{t,T}f^{''}(y_{t,T})}{f^{'}(y_{t,T})}{\sigma}_{y}^{2}(t,T)dt+ {\sigma}_{y}(t,T)dW_{t}\label{eq:YieldDyn}
\end{eqnarray}

\textbf{Proof}\\
From equation \textbf{(\ref{eq:Bond})} or \textbf{(\ref{eq:BondCMT})} above, we can write that
\begin{eqnarray}
y_{t,T} & = & f^{-1}(B_{t,T}) \nonumber\\
				&=& g(B_{t,T}) \nonumber
\end{eqnarray}

Using Ito'o lemma, we thus get

$$dy_{t,T}=g^{'}(B_{t,T})dB_{t,T}+\frac{1}{2} g^{''}(B_{t,T})d{<}B_{.,T};B_{.,T}{>}_t$$

Using equation \textbf{(\ref{eq:BondFwrd})}, we get that
\begin{eqnarray}
\frac{dy_{t,T}}{y_{t,T}}  =  \frac{1}{2} \frac{g^{''}(B_{t,T})}{g(B_{t,T})}B_{t,T}^{2}{\sigma}_{B}^{2}(t,T)dt +\frac{B_{t,T}g^{'}(B_{t,T})}{g(B_{t,T})}{\sigma}_{B}(t,T)dW_t \label{eq:YieldProof}
\end{eqnarray}

We remind that
$$g(b)=f^{-1}(b)$$
$$g^{'}(b)=\frac{1}{f^{'}(f^{-1}(b))}$$
$$g^{''}(b)=-\frac{f^{''}(f^{-1}(b))}{{\left[f^{'}(f^{-1}(b))\right]}^{3}}$$

Replacing in equation \textbf{(\ref{eq:YieldProof})}, we get the two results: $\boxdot$
\begin{eqnarray}
\frac{dy_{t,T}}{y_{t,T}}  =  -\frac{1}{2} \frac{y_{t,T}f^{''}(y_{t,T})}{f^{'}(y_{t,T})}\left(\frac{f(y_{t,T})}{y_{t,T}f^{'}(y_{t,T})}{\sigma}_{B}(t,T)\right)^{2}dt +\frac{f(y_{t,T})}{y_{t,T}f^{'}(y_{t,T})}{\sigma}_{B}(t,T)dW_t \nonumber
\end{eqnarray}\\\\
We remind that our objective is to be able to diffuse the forward yield to maturity. At this stage, we have written the yield's dynamics, but our goal is not yet reached, since we don't have the volatility of the yield, and also we don't have the initial value of the forward yield. \\

\textbf{Proposition 2}\\
\textit{The initial value of the forward yield to maturity is the following:}

$$y_{0,T}=\left(\frac{1-S(0,T,T+\theta)DF(0,T,T+\theta)-R*\Gamma(0,T,T+\theta)}{\sum_{i=1}^{{\kappa}\theta}{S(0,T,T_{i})DF(0,T,T_{i})}}+1\right)^{\kappa}-1$$

\textit{With}
\begin{eqnarray}
\Gamma(t,T,U)&=&\int_{T}^{U}{P_{t,T,u} d_{u}(1-S(t,T,u))} \nonumber\\
				&=&\int_{T}^{U}{P_{t,T,u}S(t,T,u){\lambda}(t,T,u)du} \nonumber
\end{eqnarray}

\textbf{Proof}\\
Given the initial rate curve, the initial default probability distribution and the recovery rate of the issuer, one can express the initial value of any constant maturity forward Bond as following:

$$B_{0,T}=S(0,T,T+\theta)DF(0,T,T+\theta)+\sum_{i=1}^{{\kappa}\theta}{\tilde{c}(T_{i}-T_{i-1})S(0,T,T_{i})DF(0,T,T_{i})}+R*\Gamma(0,T,T+\theta)$$

In particular, for the $CMT\theta$ related Bond, we have that (equation \textbf{(\ref{eq:CMT})}) $\tilde{c}=\kappa\left((1+y_{T,T})^{\frac{1}{\kappa}}-1\right)$; and by definition this bond should be at par. We thus find the following equality (with $(T_{i}-T_{i-1})\cong\frac{1}{\kappa}$):

$$S(0,T,T+\theta)DF(0,T,T+\theta)+\left((1+y_{T,T})^{\frac{1}{\kappa}}-1\right)\sum_{i=1}^{{\kappa}\theta}{S(0,T,T_{i})DF(0,T,T_{i})}+R*\Gamma(0,T,T+\theta)=1$$

And we have the result. $\boxdot$
\\\\
$y_{0,T}$  is a first good approximation of the forward yield. It is the value without the correction due to the convexity (convexity adjustment).

\section{Calculation of the volatility of the forward Bond}
In this section we calculate the volatility of the forward bond, when the volatility of the zero coupon bond is known. And also we suppose a deterministic Hazard rate function, with respect to the time between today and the expiry (forward start) date. The bond's volatility will be used for computing the volatility of the forward yield to maturity, through equation \textbf{(\ref{eq:VolYield})} above.\\

\textbf{Proposition 3}\\
\textit{Under the double conditions that the volatility of the zero-coupon bond is known, and that the Hazard Rate Function is deterministic with respect to the time between today and the forward start date, we have the following expression for the volatility of the constant maturity forward Bond:}
\begin{eqnarray}
\sigma_{B}(t,T)=\frac{1}{f(y_{t,T})}\textbf{(}e^{-\int_{T}^{T+\theta}{{\lambda}(t,T,v)dv}} P_{t,T,T+\theta}{\sigma}_{P}(t,T,T+\theta) \label{eq:VolBond}\\ 
 +\frac{\tilde{c}}{\kappa}\sum_{i=1}^{{\kappa}\theta}{\left[e^{-\int_{T}^{T_{i}}{{\lambda}(t,T,v)dv}} P_{t,T,T_{i}}{\sigma}_{P}(t,T,T_{i})\right]} \nonumber\\ 
 +R\int_{T}^{T+\theta}{\left[e^{-\int_{T}^{u}{{\lambda}(t,T,v)dv}}P_{t,T,u}{\sigma}_{P}(t,T,u){\lambda}(t,T,u)\right]du}\textbf{)}\nonumber
\end{eqnarray}\\

\textit{And $\lambda$ is the solution of the following pde:}
\begin{eqnarray}
R\int_{T}^{T+\theta}{\left({\lambda}(t,T,u)\left[-\int_{T}^{u}{\frac{\partial{\lambda}}{{\partial}t}(t,T,v)dv} + {\xi}(t,T,u)\right]+\frac{\partial{\lambda}}{{\partial}t}(t,T,u)\right)P_{t,T,u}e^{-\int_{T}^{u}{{\lambda}(t,T,v)dv}}du} \nonumber\\
\left(-\int_{T}^{T+\theta}{\frac{\partial{\lambda}}{{\partial}t}(t,T,v)dv} + {\xi}(t,T,T+\theta)\right)P_{t,T,T+\theta}e^{-\int_{T}^{T+\theta}{{\lambda}(t,T,v)dv}} \label{eq:hrf}\\
+\frac{\tilde{c}}{\kappa}\sum_{i=1}^{{\kappa}\theta}{\left(-\int_{T}^{T_{i}}{\frac{\partial{\lambda}}{{\partial}t}(t,T,v)dv} + {\xi}(t,T,T_{i})\right)P_{t,T,T_{i}}e^{-\int_{T}^{T_{i}}{{\lambda}(t,T,v)dv}}} =0 \nonumber
\end{eqnarray}
\newpage
\textbf{Proof}\\
Given the forward survival probability distribution and the forward zero coupon bonds, the expression of the constant maturity forward Bond is the following:
$$B_{t,T}=S(t,T,T+\theta)P_{t,T,T+\theta}+\frac{\tilde{c}}{\kappa}\sum_{i=1}^{{\kappa}\theta}{S(t,T,T_{i})P_{t,T,T_{i}}}+R*\Gamma(t,T,T+\theta)$$

Differentiating with respect to the first variable ($t$), we get 
\begin{eqnarray}
dB_{t,T} = \textbf{(}e^{-\int_{T}^{T+\theta}{{\lambda}(t,T,v)dv}} P_{t,T,T+\theta}{\sigma}_{P}(t,T,T+\theta) %\nonumber\\ 
 +\frac{\tilde{c}}{\kappa}\sum_{i=1}^{{\kappa}\theta}{\left[e^{-\int_{T}^{T_{i}}{{\lambda}(t,T,v)dv}} P_{t,T,T_{i}}{\sigma}_{P}(t,T,T_{i})\right]} \nonumber\\ 
 +R\int_{T}^{T+\theta}{\left[e^{-\int_{T}^{u}{{\lambda}(t,T,v)dv}}P_{t,T,u}{\sigma}_{P}(t,T,u){\lambda}(t,T,u)\right]du}\textbf{)}dW_{t}\nonumber\\
\textbf{(}R\int_{T}^{T+\theta}{\left({\lambda}(t,T,u)\left[-\int_{T}^{u}{\frac{\partial{\lambda}}{{\partial}t}(t,T,v)dv} + {\xi}(t,T,u)\right]+\frac{\partial{\lambda}}{{\partial}t}(t,T,u)\right)P_{t,T,u}e^{-\int_{T}^{u}{{\lambda}(t,T,v)dv}}du} \nonumber\\
\left(-\int_{T}^{T+\theta}{\frac{\partial{\lambda}}{{\partial}t}(t,T,v)dv} + {\xi}(t,T,T+\theta)\right)P_{t,T,T+\theta}e^{-\int_{T}^{T+\theta}{{\lambda}(t,T,v)dv}} \nonumber\\
+\frac{\tilde{c}}{\kappa}\sum_{i=1}^{{\kappa}\theta}{\left(-\int_{T}^{T_{i}}{\frac{\partial{\lambda}}{{\partial}t}(t,T,v)dv} + {\xi}(t,T,T_{i})\right)P_{t,T,T_{i}}e^{-\int_{T}^{T_{i}}{{\lambda}(t,T,v)dv}}}\textbf{)}dt\nonumber
\end{eqnarray}

with

$$S(t,T,U)=e^{-\int_{T}^{U}{{\lambda}(t,T,v)dv}}$$

Then the volatility of the forward bond is the Brownian motion coefficient divided by the value of the forward bond.

Since the forward bond should be a local martingale, the drift term of the expression above should vanish. We thus get the pde, and this ends the proof.  $\boxdot$\\\\

Looking at equation \textbf{(\ref{eq:hrf})}, we see that the hazard rate function and thus the survival or default probability distribution of the bond issuer depends on the coupon rate of the Bond. This is a limitation of this framework, and this observation shows that the hypothesis of a deterministic hazard rate function is obviously wrong. To be perfect one should use a stochastic hazard rate function, and this will include the volatility of the time-to-default, and the correlation between the time-to-default and the interest rate.

Now we do have the dynamics of the forward yield; we have the initial value of the forward yield; we have the volatility of the forward bond and the volatility of the forward yield through equation \textbf{(\ref{eq:VolYield})}. We can thus diffuse the forward yield to maturity and price any derivative on this underlying.

However, it is not easy to solve equation \textbf{(\ref{eq:hrf})}, and when dealing with the constant maturity treasury, the coupon rate is a function of the terminal value of the yield, and we don't have it. 

\section{Simplification of the hazard rate function and proposed algorithm for diffusing the yield when dealing with the CMT}
In this section we make the assumption that the Hazard rate function does not depend on the time between the expiry (forward start date) and the bond's maturity. The coupon rate here is the $CMT\theta$.
In this context we propose a numerical solution for equation \textbf{(\ref{eq:hrf})}, and finally we propose a consistent algorithm for diffusing the forward yield, and to price, in Monte Carlo, any pay off on the $CMT\theta$
\\\\
We actually write that
$${\lambda}(t,T,U)={\lambda}(t,T)$$
Then we have
$$S(t,T,U)=e^{-(U-T){\lambda}(t,T)}$$ 
Equation \textbf{(\ref{eq:hrf})} becomes:
\begin{eqnarray}
R\int_{T}^{T+\theta}{\left({\lambda}(t,T)\left[-(u-T)\frac{\partial{\lambda}}{{\partial}t}(t,T) + {\xi}(t,T,u)\right]+\frac{\partial{\lambda}}{{\partial}t}(t,T)\right)P_{t,T,u}e^{-(u-T){\lambda}(t,T)}du} \nonumber\\
\left(-{\theta}\frac{\partial{\lambda}}{{\partial}t}(t,T) + {\xi}(t,T,T+\theta)\right)P_{t,T,T+\theta}e^{-{\theta}{\lambda}(t,T)} \nonumber\\
+\left((1+y_{T,T})^{\frac{1}{\kappa}}-1\right)\sum_{i=1}^{{\kappa}\theta}{\left(-(T_{i}-T)\frac{\partial{\lambda}}{{\partial}t}(t,T) + {\xi}(t,T,T_{i})\right)P_{t,T,T_{i}}e^{-(T_{i}-T){\lambda}(t,T)}} =0 \nonumber
\end{eqnarray}

Still the pde is not easy to integrate. In the next lines we propose a numerical resolution. Given a discrete subdivision of the time space between 0 and T, we use the following approximation for the first derivative:
$$\frac{\partial{\lambda}}{{\partial}t}(t_{j},T) \cong \frac{{\lambda}(t_{j+1},T)-{\lambda}(t_{j-1},T)}{2{\Delta}t_{j}} \mbox{  ;   } 0 < t_{j} \leq T$$

We input this in the pde and we get the following expression for the hazard rate function at each time step:
\begin{eqnarray}
{\lambda}(t_{j+1},T) = \frac{\Phi(t_{j-1},t_{j},T)}{\Psi(t_{j},T)} \label{eq:discHRF}
\end{eqnarray}

With
\begin{eqnarray}
\Phi(t_{j-1},t_{j},T)=\left(\frac{\theta}{2}{\lambda}(t_{j-1},T)  + {\Delta}t_{j}{\xi}(t_{j},T,T+\theta)\right)e^{-{\theta}{\lambda}(t_{j},T)}P_{t_{j},T,T+\theta} \nonumber\\
+\left((1+y_{T,T})^{\frac{1}{\kappa}}-1\right)\sum_{i=1}^{{\kappa}\theta}{\left(\frac{(T_{i}-T)}{2}{\lambda}(t_{j-1},T)  + {\Delta}t_{j}{\xi}(t_{j},T,T_{i})\right)e^{-(T_{i}-T){\lambda}(t_{j},T)}P_{t_{j},T,T_{i}}} \nonumber\\
R\int_{T}^{T+\theta}{\left({\lambda}(t_{j},T)\left[\frac{(u-T)}{2}{\lambda}(t_{j-1},T) + {\Delta}t_{j}{\xi}(t_{j},T,u)\right]-\frac{{\lambda}(t_{j-1},T)}{2}\right)e^{-(u-T){\lambda}(t_{j},T)}P_{t_{j},T,u}du} \nonumber
\end{eqnarray}

\begin{eqnarray}
\Psi(t_{j},T)=\frac{\theta}{2}e^{-{\theta}{\lambda}(t_{j},T)}P_{t_{j},T,T+\theta} %\nonumber\\
+\left((1+y_{T,T})^{\frac{1}{\kappa}}-1\right)\sum_{i=1}^{{\kappa}\theta}{\frac{(T_{i}-T)}{2}e^{-(T_{i}-T){\lambda}(t_{j},T)}P_{t_{j},T,T_{i}}} \nonumber\\
\frac{R}{2}\int_{T}^{T+\theta}{\left[1+(u-T){\lambda}(t_{j},T)\right]e^{-(u-T){\lambda}(t_{j},T)}P_{t_{j},T,u}du} \nonumber
\end{eqnarray}

\begin{itemize}
	\item{${\lambda}(t_{0},T) = -\frac{1}{\theta}\ln\left(\frac{S(t_{0},T+\theta)}{S(t_{0},T)}\right)$ is an estimation of the initial forward ha\-zard rate;}
	\item{${\lambda}(t_{-1},T) =0$;}
	\item{${\Delta}t_{j} = t_{j}-t_{j-1}$.}
\end{itemize}

We have everything now to propose the final algorithm for the pricing.\\\\

\textbf{Algorithm}

\begin{enumerate}
	\item{Compute the initial forward yield $y_{0,T}$, as specified in the \textbf{proposition 2} above;}
\end{enumerate}
	At each time step: $0 < t_{j} \leq T$
\begin{enumerate}
	\item[2]{Diffuse equation \textbf{(\ref{eq:ZCBondFwrd})}, for all maturities from $T$ to $T+\theta$: $T \leq u \leq T+\theta$;}
	\item[3]{Compute the Hazard rate function, and the survival probability as in expression \textbf{(\ref{eq:hrf})}. Use $y_{t_{j-1},T}$ for computing the coupon rate: $y_{T,T} \cong y_{t_{j-1},T}$;}
	\item[4]{Compute the volatility of the forward bond $\sigma_{B}(t_{j},T)$, using equation \textbf{(\ref{eq:VolBond})}. Use $y_{t_{j-1},T}$ for computing the coupon rate: $y_{T,T} \cong y_{t_{j-1},T}$;}
	\item[5]{Compute the volatility of the yield using equation \textbf{(\ref{eq:VolYield})}: 
	$${\sigma}_{y}(t_{j},T)=\frac{f(y_{t_{j-1},T})}{y_{t_{j-1},T}f^{'}(y_{t_{j-1},T})}{\sigma}_{B}(t_{j-1},T) $$}
	\item[6]{Diffuse the yield to maturity using equation \textbf{(\ref{eq:YieldDyn})}, and move to the next step;}
	\item[7]{At maturity calculate de CMT using equation \textbf{(\ref{eq:CMT})}.}
\end{enumerate}

\section{Application and results in Hull and White Model}
\textbf{Hull and White model}\\
In this model, one makes the assumption that the short term interest rate is a mean reverting, normal distributed process:
$$dr_{t}=\alpha(\tilde{r}-r_{t})dt+{\sigma}dW_t$$

\begin{itemize}
	\item{$\sigma$ is the volatility parameter. It is supposed to be constant;} 
	\item{$\alpha$ is the mean reversion parameter. It characterizes the speed of reversion of the interest rate toward  $\tilde{r}$. the bigger is $\alpha$, the faster the short rate reverts towards $\tilde{r}$. It is supposed to be constant.}
	\item{$\tilde{r}$ is the level toward reverts the short interest rate;}
\end{itemize}

$\tilde{r}$ could be deterministic or stochastic(2 factors). When $\tilde{r}$ is not stochastic, then the volatility of a zero coupon Bond with maturity $T$ is given by:
$$\sigma_{P}(t,T)={\sigma}\frac{(1-e^{-{\alpha}(T-t)})}{\alpha}$$
And the volatility of the forward zero coupon Bond is
$$\sigma_{P}(t,T,U)=\frac{\sigma}{\alpha}(e^{-{\alpha}(T-t)}-e^{-{\alpha}(U-t)})$$
Once we have calibrated Hull and White model parameters on market ins\-trument prices, we can calculate the CMT using the algorithm described in the previous section; and finally we can price derivatives on the $CMT\theta$.\\\\

\textbf{Tests and results}\\
All the tests below have been done in JPY currency.  The initial hazard rate function (piece-wise constant) has been stripped from the spot prices of the Japanese Government Bonds (JGB). The tests have been realized on the same date: 28/03/2012. The rolling constant maturity is ten years: $\theta=10$
\begin{figure}[htbp]
	\centering
		\includegraphics[height = 5.1cm, width = 13cm]{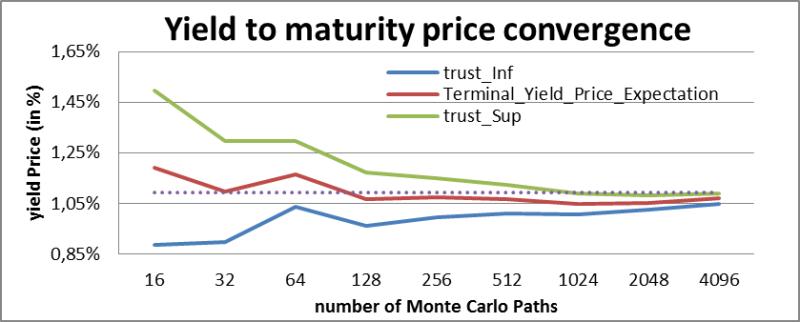}
	\caption{\textit{test of convergence. 1 year expiry: 28/03/2013; $R=20\%$; discretization step = 1 day; $\alpha=10\%$; $\sigma=1\%$. This figure shows that the price of the terminal yield converges very quickly: from 512 paths we have satisfactory price.}}
	\label{fig:conv1}
\end{figure}

\begin{figure}[htbp]
	\centering
		\includegraphics[height = 5.1cm, width = 13cm]{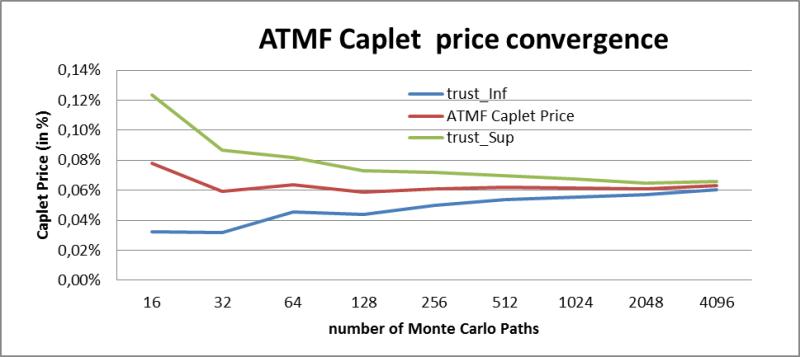}
	\caption{\textit{test of convergence. 1 year expiry: 28/03/2013; $R=20\%$; discretization step = 1 day; $\alpha=10\%$; $\sigma=1\%$. This figure shows a rapid convergence for the ATMF (at the money forward) Caplet price: from 128 paths.}}
	\label{fig:conv2}
\end{figure}

\begin{figure}[htbp]
	\centering
		\includegraphics[height = 7cm, width = 13cm]{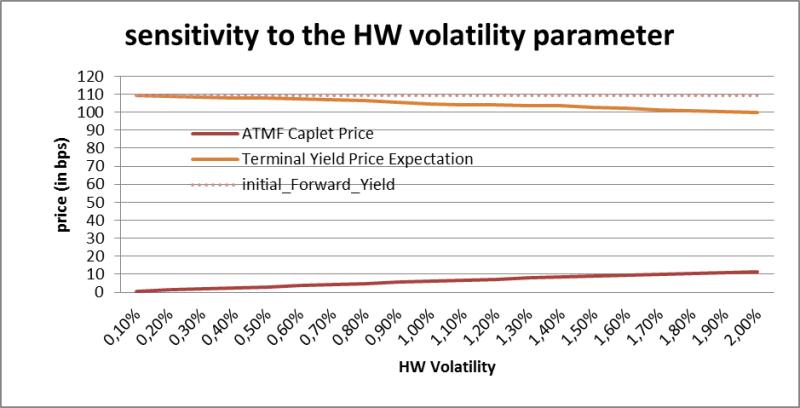}
	\caption{\textit{sensitivity to model Parameters. Expiry= 28/03/2013; Maturity=23/06/2013; $R=20\%$; discretization step = 1 day; $\alpha=10\%$; Mtcl Paths=1024. On this figure we see that the expectation of the terminal yield decreases  with the short rate volatility parameter, whereas the ATMF caplet price increases.}}
	\label{fig:sensiParamVol}
\end{figure}

\begin{figure}[htbp]
	\centering
		\includegraphics[height = 7cm, width = 13cm]{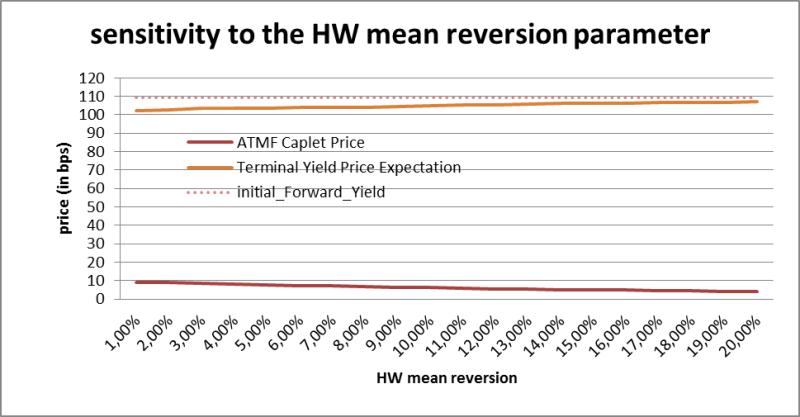}
	\caption{\textit{sensitivity to model Parameters. Expiry= 28/03/2013; Maturity(Caplet)=23/06/2013; $R=20\%$; discretization step = 1 day; $\sigma=1\%$; Mtcl Paths=1024. On this figure we see that the expectation of the terminal yield increases with the short rate volatility parameter, whereas the ATMF caplet price decreases.}}
	\label{fig:sensiParamRev}
\end{figure}

\begin{figure}[htbp]
	\centering
		\includegraphics[height = 6.5cm, width = 13cm]{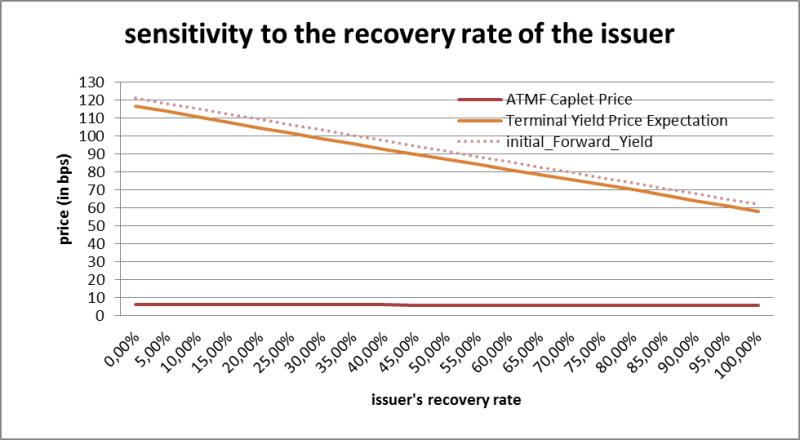}
	\caption{\textit{sensitivity to Recovery rate. Expiry= 28/03/2013; Maturity (Caplet)=23/06/2013; discretization step = 1 day; $\sigma=1\%$; $\alpha=10\%$; Mtcl Paths=1024. On this figure we see that the expectation of the terminal yield decreases with the recovery rate, whereas the ATMF caplet price is almost constant with the recovery rate.}}
	\label{fig:sensiParamRecov}
\end{figure}

\begin{figure}[htbp]
	\centering
		\includegraphics[height = 6.5cm, width = 13cm]{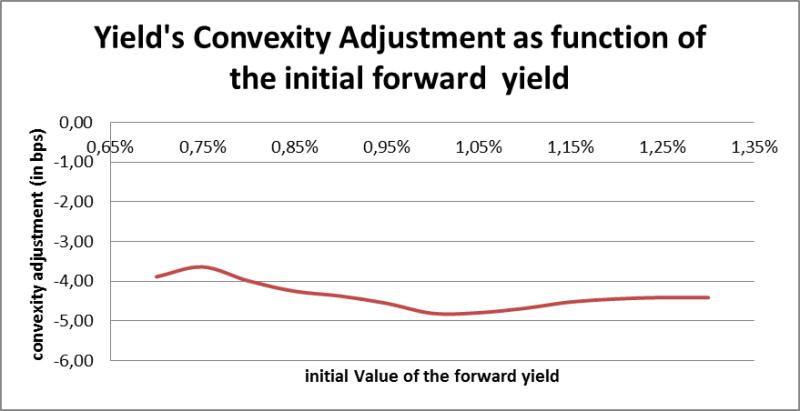}
	\caption{\textit{sensitivity of the convexity adjustment to the initial value of the forward yield. The convexity adjustment being the difference between the expectation of the terminal yield, and the initial value of the forward yield. Expiry= 28/03/2017; discretization step = 1 day; $\sigma=1\%$; $\alpha=10\%$; $R=20\%$; Mtcl Paths=1024. On this figure we see that the convexity adjustment is almost constant with respect to the initial forward yield.}}
	\label{fig:ConvAdj}
\end{figure}

\begin{figure}[htbp]
	\centering
		\includegraphics[height = 7cm, width = 13cm]{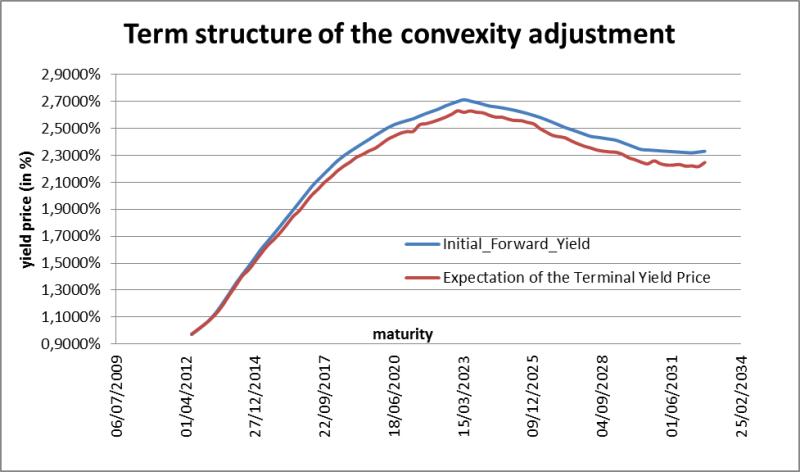}
	\caption{\textit{ discretization step = 1 day; $\sigma=1\%$; $\alpha=10\%$; Mtcl Paths=4096; $R=20\%$.We observe on this figure that the convexity adjustment increases with the maturity}}
	\label{fig:convAdjStruct}
\end{figure}

\begin{figure}[htbp]
	\centering
		\includegraphics[height = 7cm, width = 13cm]{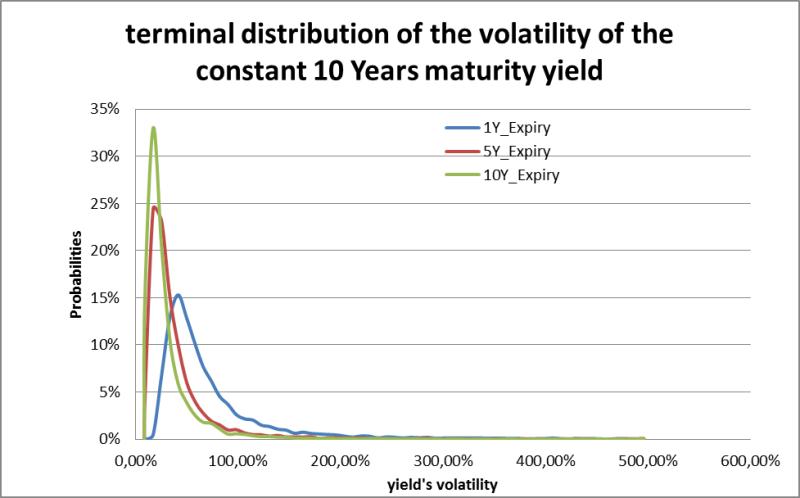}
	\caption{\textit{discretization step = 1 day; $\sigma=1\%$; $\alpha=10\%$; Mtcl Paths=8192; $R=20\%$.We observe on this figure that the volatility of the yield to maturity is log normal, or very close to a such distribution.}}
	\label{fig:distVol}
\end{figure}

\begin{table}
	\centering
		\begin{tabular}{|c|c|c|c|}
			\hline
			 & \textbf{1 Year Expiry} & \textbf{5 Years Expiry} & \textbf{10 Years Expiry}\\
			\hline
			\textbf{Min} &	$19,21\%$	 & $10,00\%$ & $9,09\%$\\
			\hline
			\textbf{Max}	& $493,61\%$ &	$495,10\%$ &	$460,21\%$\\
			\hline
			\textbf{Average} &	$77,78\%$ &	$43,70\%$ &	$35,08\%$\\
			\hline
			\textbf{Standard deviation} &	$60,28\%$	& $39,48\%$	& $31,06\%$\\
			\hline
			\textbf{skewness} &	$309,78\%$ &	$493,99\%$ &	$485,28\%$\\
			\hline
			\textbf{kurtosis} &	$1220,45\%$ &	$3588,85\%$ &	$3672,24\%$\\
			\hline
		\end{tabular}
	\caption{Statistical data for the distribution of the yield to maturity's volati\-lity}
	\label{tab:stat1}
\end{table}

\begin{figure}[htbp]
	\centering
		\includegraphics[height = 7.5cm, width = 13cm]{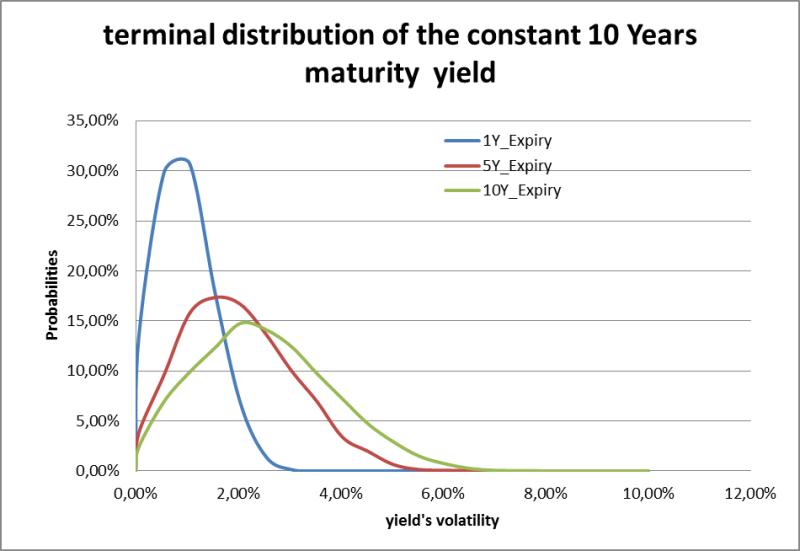}
	\caption{\textit{discretization step = 1 day; $\sigma=1\%$; $\alpha=10\%$; Mtcl Paths=8192; $R=20\%$.We observe on this figure that the yield to maturity is not log normal.}}
	\label{fig:distYield}
\end{figure}

\begin{table*}
	\centering
		\begin{tabular}{|c|c|c|c|}
			\hline
			 & \textbf{1 Year Expiry} & \textbf{5 Years Expiry} & \textbf{10 Years Expiry}\\
			\hline
			\textbf{Min} &	$0,07\%$	 & $0,10\%$ & $0,12\%$\\
			\hline
			\textbf{Max}	& $3,76\%$ &	$8,03\%$ &	$8,70\%$\\
			\hline
			\textbf{Average} &	$1,20\%$ &	$2,25\%$ &	$2,77\%$\\
			\hline
			\textbf{Standard deviation} &	$0,57\%$	& $1,08\%$	& $1,31\%$\\
			\hline
			\textbf{skewness} &	$45,25\%$ &	$48,22\%$ &	$40,34\%$\\
			\hline
			\textbf{kurtosis} &	$-5,80\%$ &	$-4,87\%$ &	$-16,46\%$\\
			\hline
		\end{tabular}
	\caption{Statistical data for the distribution of the yield to maturity}
	\label{tab:stat2}
\end{table*}

\begin{figure}[htbp]
	\centering
		\includegraphics[height = 7.5cm, width = 13cm]{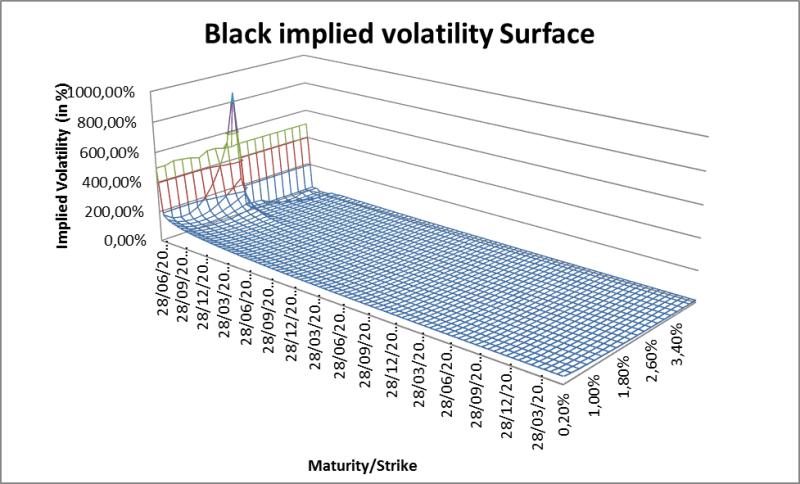}
	\caption{\textit{Black implied volatility surface, related to Caps and Floors prices on the CMT10. Caplets or Floorlets are paid with a frequency equal to 3 Months.  Discretization step = 1 day;  $\sigma=1\%$; $\alpha=10\%$; Mtcl Paths=1024; $R=20\%$.}}
	\label{fig:BlackSurf}
\end{figure}

\begin{figure}[htbp]
	\centering
		\includegraphics[height = 7.5cm, width = 13cm]{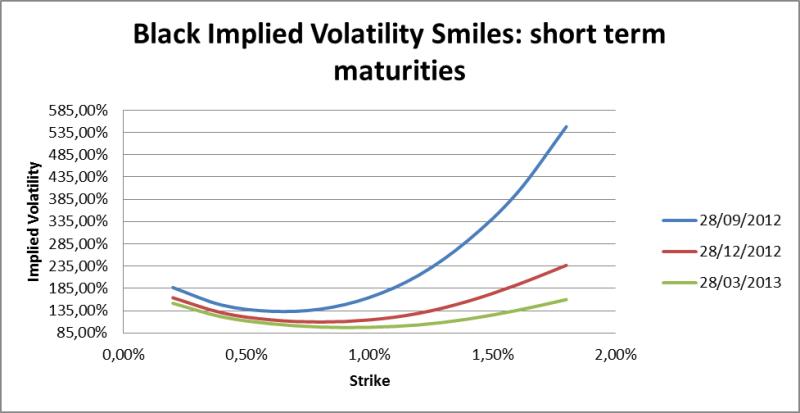}
	\caption{\textit{Short term maturities Black implied volatility smiles, related to Caps and Floors prices on the CMT10. Caplets or Floorlets are paid with a frequency equal to 3 Months. Discretization step = 1 day; $\sigma=1\%$; $\alpha=10\%$; Mtcl Paths=1024; $R=20\%$. The curve is very important on the short term implied volatility.}}
	\label{fig:BlackSmileShort}
\end{figure}

\begin{figure}[htbp]
	\centering
		\includegraphics[height = 7.5cm, width = 13cm]{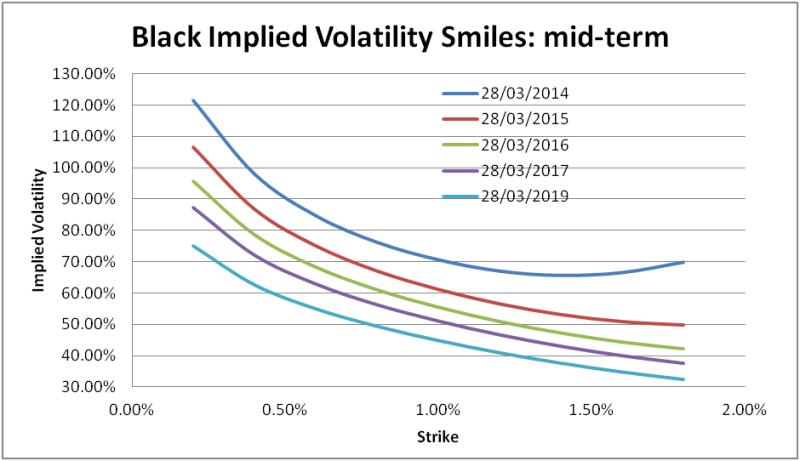}
	\caption{\textit{Mid-term maturities Black implied volatility smiles, related to Caps and Floors prices on the CMT10. Caplets or Floorlets are paid with a frequency equal to 3 Months. Discretization step = 1 day;$\sigma=1\%$; $\alpha=10\%$; Mtcl Paths=1024;  $R=20\%$. }}
	\label{fig:BlackSmileMid}
\end{figure}

\begin{figure}[htbp]
	\centering
		\includegraphics[height = 7.5cm, width = 13cm]{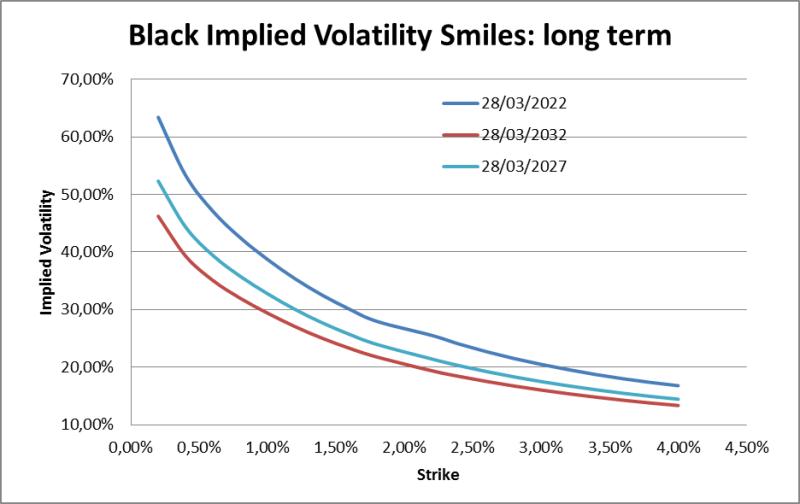}
	\caption{\textit{Long-term maturities Black implied volatility smiles, related to Caps and Floors prices on the CMT10. Caplets or Floorlets are paid with a frequency equal to 3 Months. Discretization step = 1 day;$\sigma=1\%$; $\alpha=10\%$; Mtcl Paths=1024;  $R=20\%$. }}
	\label{fig:BlackSmileLong}
\end{figure}

\section*{Conclusion}
In this paper we have used the martingale results to derive the dynamics of the forward yield to maturity, and to calculate the volatility of the forward yield to maturity. This is possible when we have the volatility of the zero-coupon Bond. We made the assumption of a deterministic hazard rate function, with respect to the time between today and the expiry date. We have proposed a result for the initial value of the forward yield to maturity.
These results enable us to diffuse the yield to maturity, and to price derivatives on the forward CMT or the forward bonds.\\

As application we have supposed a Hull and white model for the short interest rate, we have computed the volatility of the zero-coupon bonds, then we have calculated the convexity adjustment on the yield to maturity, we have calculated the terminal yield to maturity distribution, the distribution of the terminal volatility of the yield to maturity, and Black implied volati\-lities surface related to Caps and Floors on the CMT10.\\

As results we observe that the convexity adjustment of the yield increases with the maturity. We observe that the volatility of the yield to maturity seems to be log normal, whereas the yield to maturity itself is not log normal distributed. The black implied volatility generated is decreasing with respect to the maturity, for any fixed strike. The implied volatility smile has an important curve for short maturities (until 1 Year maturity), and then the curve decreases as maturity goes higher.\\
 
As extension to this work, one should release the hypothesis of determi\-nistic hazard rate function. This will includes the volatility of the Hazard rate function, and the correlation between the time to default and the interest rate.

\newpage
\section*{References}
Benhamou E.: 2000, A Martingale Result for the Convexity Adjustment in the Black Pricing Model, \textit{London School of Economics}, Working Paper. March.\\
\\
Benhamou E.: 2000, Pricing Convexity Adjustment with Wiener Chaos, \textit{London School of Economics}, Discussion Paper 351. April.\\
\\
Brigo D. and Mercurio F.: 2005, Interest Rate Models - Theory and Practice: With Smile, Inflation and Credit, \textit{Springer Finance}. Second Edition\\
\\
Jordan J. V. and Mansi S. A.: 2000, How Well do constant-maturity treasuries approximate the on-the-run Term structure?, \textit{The Journal of Fixed Income}. September.\\
\\
Kouokap Youmbi D. : 2010, Pricing of CDS, Bond and CDO, \textit{Working Note}, Societe Generale.\\
\\
Press W. H., Teukolsky S. A., Vetterling W. T., Flannery B. P.:  Numerical Recipes in C, \textit{Cambridge University Press}, pp. 147-161, Second Edition.

\end{document}